\documentclass[letterpaper,10pt,conference]{IEEEtran}
\IEEEoverridecommandlockouts
\usepackage{ifpdf}
\ifpdf
    \usepackage[pdftex]{graphicx}
    \usepackage[update]{epstopdf}
\else
	\usepackage{graphicx}
\fi
\usepackage{wrapfig,bbm}
\usepackage{enumitem,color}
\usepackage{amsthm,multirow}

\usepackage[ruled,vlined]{algorithm2e}

\usepackage[mathscr]{euscript}
\usepackage{eufrak}
\usepackage{bbold}
\usepackage{array}
\usepackage{amsmath,amsthm}
\usepackage{amssymb}
\usepackage{amstext}
\usepackage{cite,setspace}
\usepackage{algorithm2e}

\DeclareGraphicsExtensions{.pdf,.png,.jpg,.eps}

\begin{document}

\setlength{\columnsep}{0.21in}

\title{\huge A New Approach to Compute Information Theoretic Outer Bounds and Its Application to Regenerating Codes\thanks{This work was supported in part by the National Science Foundation via Grants CCF-18-16546 and CCF-20-07067.}}

\author{
\IEEEauthorblockN{Wenjing Chen and Chao Tian}
\IEEEauthorblockA{Department of Electrical and Computer Engineering\\
Texas A\&M University\\
{\sffamily  \{jj9754,chao.tian\}@tamu.edu}}
}

\textfloatsep=0.25cm

\maketitle

\begin{abstract}
The study of the fundamental limits of information systems is a central theme in information theory. Both the traditional analytical approach and the recently proposed computational approach have significant limitations, where the former is mainly due to its reliance on human ingenuity, and the latter due to its exponential memory and computational complexity. In this work, we propose a new computational approach to tackle the problem with much lower memory and computational requirements, which can naturally utilize certain intuitions, but also can maintain the strong computational advantage of the existing computational approach. A reformulation of the underlying optimization problem is first proposed, which converts the large linear program to a maximin problem. This leads to an iterative solving procedure, which uses the LP dual to carry over learned evidence between iterations. The key in the reformulated problem is the selection of good information inequalities, with which a relaxed LP can be formed. A particularly powerful intuition is a potentially optimal code construction, and we provide a method that directly utilizes it in the new algorithm. As an application, we derive a tighter outer bound for the storage-repair tradeoff for the $(6,5,5)$ regenerating code problem, which involves at least 30 random variables and is impossible to compute with the previously known computational approach.
\end{abstract}

\section{Introduction}

The study of the fundamental limits of information systems is a central theme in information theory since its invention by Claude Shannon in 1948 \cite{Shannon:48}.  
There have been many advances in this area over the years \cite{CoverThomas,el2011network,yeung2006network,yeung2008information}, and more and more information systems have been studied using an information theoretic approach, such as coding for distributed data storage \cite{Dimakis:10,RashmiShah:11}, coded caching \cite{MaddahAliNiesen:14}, private information retrieval \cite{sun2017PIRcapacity}, and straggler-resilient coded computation \cite{lee2017speeding}.

Fundamental limits of information systems, i.e., outer bounds or converse bounds, have traditionally been derived using an analytical approach, which are usually presented as a sequence of bounding steps involving information inequalities. This approach requires a deep understanding on the problem under consideration, familiarity with information theoretic techniques, and perhaps most importantly, a heavy dose of human ingenuity. As information systems become more complex, this approach becomes rather unwieldy. Recently, a computational approach has been proposed to address such difficulties, and a few notable results have been obtained with the assistance of this approach \cite{Tian:JSAC13,Tian:15-2,tian2018symmetry,shao2020fundamental,TianLiu:15,Walsh:16,tian2018shannon,tian2020storage,guo2021new,li2020conditional}. 

Despite the successes of the computational approach in these studies, its limitation has become increasingly apparent. The main idea of this approach is that the class of information inequalities usually used in such problems (i.e., Shannon-type inequalities) are linear in terms of the joint entropies \cite{Yeung:97}; by viewing the derivation of the converse bound as an optimization problem with the joint entropies as variables, and the information inequalities and problem-specific conditions as constraints in a linear program (LP), outer bounds and the corresponding proofs can be found by solving the LP and the dual. The key difficulty in this approach is that the number of Shannon-type information inequalities grows exponentially \cite{yeung2008information} in the total number of random variables, meaning that for problems of slightly larger scales, it is unrealistic to complete the needed computation. Although symmetry and implication relations can be used to reduce the scale of the LP, which indeed made some seemingly impossible cases become possible to compute \cite{Tian:JSAC13,tian2018symmetry,tian2021computational}, they cannot fully resolve the memory and computational hurdle \cite{ZhangTian:17TCOM}. 

In view of the difficulty discussed above, new techniques are sorely needed to break the aforementioned memory and computation barriers. For this purpose, it is beneficial to consider how human researchers approach the problem differently from the computational approach. Firstly, humans usually take a trial-and-error approach, and it may take multiple attempts to find a viable bound. Secondly,  humans are very good at using intuitions, which can be obtained from \textit{side information} (SI) channels such as potentially optimal codes, easy-to-study smaller instances, genie-aided (relaxed) systems, etc.. In contrast, the existing computational approach is largely single-shot, exhaustive (on using Shannon-type inequalities), and not relying on any SI. Therefore, we need to make the existing approach more ``intelligent'' in some way. 

In this work, we propose a new and more intelligent computational approach based on a reformulated optimization problem, which is motivated by the following critical observation. Although the number of Shannon-type inequalities can be extremely large to start with, the number of effective inequalities in the eventual proof is usually quite small (often only tens, when the LPs may have hundreds of thousands or even millions of random variables); see \cite{Tian:JSAC13,Tian:15-2,tian2018symmetry} for specific examples. A direct consequence of this observation is that if the set of effective inequalities could be identified beforehand, solving this small LP and obtaining the bound would be quite simple; if it could be approximately identified, then solving the LP would still be very fast and still yield a strong converse bound. %Moreover, we can, and will likely need to, make multiple attempts to find a good set of inequalities that leads to a good outer bound. 

This new formulation converts the problem of solving for the optimal solution of a large LP into the problem of finding a good (not necessarily optimal) solution of a discrete optimization problem (more precisely, a discrete maximin problem). The (maximization) variables in the reformulation can be viewed as the selected set of inequalities, and the objective function is the optimal LP value under this set of inequality constraints. The immediate benefit is a much lighter requirement on memory to perform pre-processing and problem setup. It also leads to a natural iterative procedure, where the LP dual can be used to carry over learned evidence between iterations. More importantly, the key to solve this optimization problem becomes finding a good subset of inequalities, for which other SI and previous experience can be used to narrow down the choices. In this work, we use potentially optimal code constructions to narrow down the choice of the information inequalities. The proposed approach is applied to the regenerating code problem \cite{Dimakis:10} with $n=6$ nodes, yielding new outer bounds  which is considerably tighter than those in the literature obtained using the analytical approach. This problem, which involves at least 30 random variables, was impossible to compute with the previous approach.

\section{The entropy linear program framework}
\label{sec:LPframework}

The entropy LP framework was first described by Yeung \cite{Yeung:97} for the purpose of proving information inequalities, and was developed further in several more recent works \cite{Tian:JSAC13,tian2018symmetry,tian2021computational} for the purpose of directly establishing outer bounds. We briefly review the general approach next for the latter purpose.

Suppose all the relevant quantities in a particular information system (i.e., a coding problem) are represented as random variables $(X_1,X_2,\ldots,X_N)$. The derivation of a fundamental limit in an information system may be understood conceptually as the following optimization problem:
\begin{align}
\text{minimize: }&\text{a weighted sum of joint entropies} \label{eqn:LPmin}\\
\text{subject to: }&\text{(I) generic information constraints} \label{eqn:contraintI}\\
&\text{(II) problem-specific information constraints}, \label{eqn:contraintII}
\end{align}
where the variables in this optimization problem are all the joint entropies on the random variables $X_1,X_2,\ldots,X_N$, as well as certain additional problem variables such as rates. Therefore, there are~$2^N-1$ variables of the form of $H(X_{\mathcal{A}})$ where $\mathcal{A}\subseteq \{1,2,\ldots,N\}$. The objective function is denoted as $f_0$. Intuitively, since any code must satisfy these constraints, the optimal value of the LP immediately provides a lower bound to the given information system.

To obtain a strong outer bound, we wish to include all the Shannon-type inequalities as generic constraints in the first group of constraints. Yeung identified a minimal set of constraints which are called elemental inequalities \cite{Yeung:97,Yeung:book}:
\begin{align}
&H(X_i|X_{\mathcal{A}})\geq 0,\quad  i\in \{1,2,\ldots,N\},\notag\\
&\qquad\qquad\qquad\qquad\qquad\mathcal{A}\subseteq \{1,2,\ldots,N\}\setminus \{i\}\label{eqn:Shannontype1}\\
&I(X_i;X_j|X_{\mathcal{A}})\geq 0, \quad i\neq j,\, i,j\in \{1,2,\ldots,N\},\notag\\
&\qquad\qquad\qquad\qquad\qquad\mathcal{A}\subseteq \{1,2,\ldots,N\}\setminus\{i,j\}.\label{eqn:Shannontype2}
\end{align}
There are $N+{N \choose 2} 2^{N-2}$ elemental inequalities, and we denote this set of inequalities as $(\text{I})$.

The second group of constraints (II) are the problem specific constraints. These are usually the implication relations required by the system (i.e., the specific coding requirements), and symmetry relations. For example, if $X_4$ is a coded representation of $X_1$ and $X_2$, then it can be represented as 
\begin{align}
H(X_4|X_1,X_2)=H(X_1,X_2,X_4)-H(X_1,X_2)=0,\label{eqn:implication}
\end{align}
which is a linear constraint. This group of constraints may also include independence and conditional independence relations. The two groups of constraints are both linear in terms of the optimization problem variables, i.e., the $2^N-1$ joint entropies and the rate variables. 

As an example, consider a hypothetical problem with 30 random variables. There will be roughly one billion variables in the LP, and over one hundred billion Shannon-type inequalities. Representing this problem and pre-processing it for symmetry and implication reductions require close to one terabyte of memory, which is impossible to complete on a standard workstation, and more importantly, solving an LP of such a scale is beyond the capability of any realistic LP solvers\footnote{Modern special-purpose large-memory server may indeed have several terabytes of RAM, however, although modern LP solvers are very powerful, they usually will have difficulty with LPs with either the number of variables or the number of constraints at the million-order scale, or on more difficult problems at a much smaller scale.}.

\section{The Proposed Alternative Computational Approach}
\label{sec:pro}

In this section, we introduce a new reformulated optimization problem, which naturally leads to an iterative optimization procedure. We further provide a heuristic procedure to select important information inequalities, and a method to utilize side information of potentially optimal code constructions.

\subsection{A Reformulated Optimization Problem}
We introduce a new optimization problem reformulation, i.e., the subset selection problem mentioned earlier. Mathematically, denote the optimal value of the problem in (\ref{eqn:LPmin}-\ref{eqn:contraintII}) as $P^*(\text{I\&II})$, where I\&II indicates that the solution is obtained under the full set of constraints; we will use a similar notation for the optimal value under different sets of constraints.

Now consider a subset $\text{I}_p\subseteq \text{I}$ of Shannon-type inequailities, and it follows that $P^*(\text{I\&II})\geq P^*(\text{I}_p\&\text{II})$. On the other hand, if there exists a subset $\text{I}^*_p\subseteq \text{I}$ with $|\text{I}^*_p|\leq \kappa$, such that $P^*(\text{I\&II})= P^*(\text{I}^*_p\&\text{II})$, then we can write 
\begin{align}
P^*(\text{I\&II}) = \max_{\text{I}_p\subseteq \text{I}:|\text{I}_p|=\kappa}P^*(\text{I}_p\&\text{II})=\max_{\text{I}_p\subseteq \text{I}:|\text{I}_p|=\kappa}\min_{\text{I}_p\&\text{II}}f_0. \label{eqn:searchoptimalsubset}
\end{align}
Thus, the original optimization problem can be converted to the problem of finding an optimal set $\text{I}_p$ on the right hand side of (\ref{eqn:searchoptimalsubset}). For small subset of cardinality $\kappa$, the inner minimization is easy to compute. Moreover, since
\begin{align}
P^*(\text{I\&II}) \geq P^*(\text{I}_p\&\text{II}),\quad \text{I}_p\subseteq \text{I},\quad |\text{I}_p|=\kappa\label{eqn:lsubset},
\end{align}
choosing any feasible subset $\text{I}_p$ yields a valid outer bound. Since the value of $\kappa$ is unknown a priori, it will be viewed as a hyper-parameter during the computation and adjusted dynamically at run-time. Recall the observation previously discussed in Section I, which suggests that this value will be much smaller than the total number of elemental Shannon-type inequalities, thus making the problem computable.

A naive algorithm naturally fits the reformulated optimization problem is exhaustive enumeration. More precisely, we can enumerate all the subset $\text{I}_p$ of a fixed cardinality, and for each such set, compute $P^*(\text{I}_p\&\text{II})$ using an LP solver; the maximum of such computed values will be the outer bound being sought after.  It is not too difficult to see that this naive approach will induce an astronomical computation cost for moderate-sized problems, and thus not practical. 

\subsection{An Iterative Optimization Procedure Using LP Dual}

\begin{figure}
\centering
\includegraphics[width=0.2\textwidth]{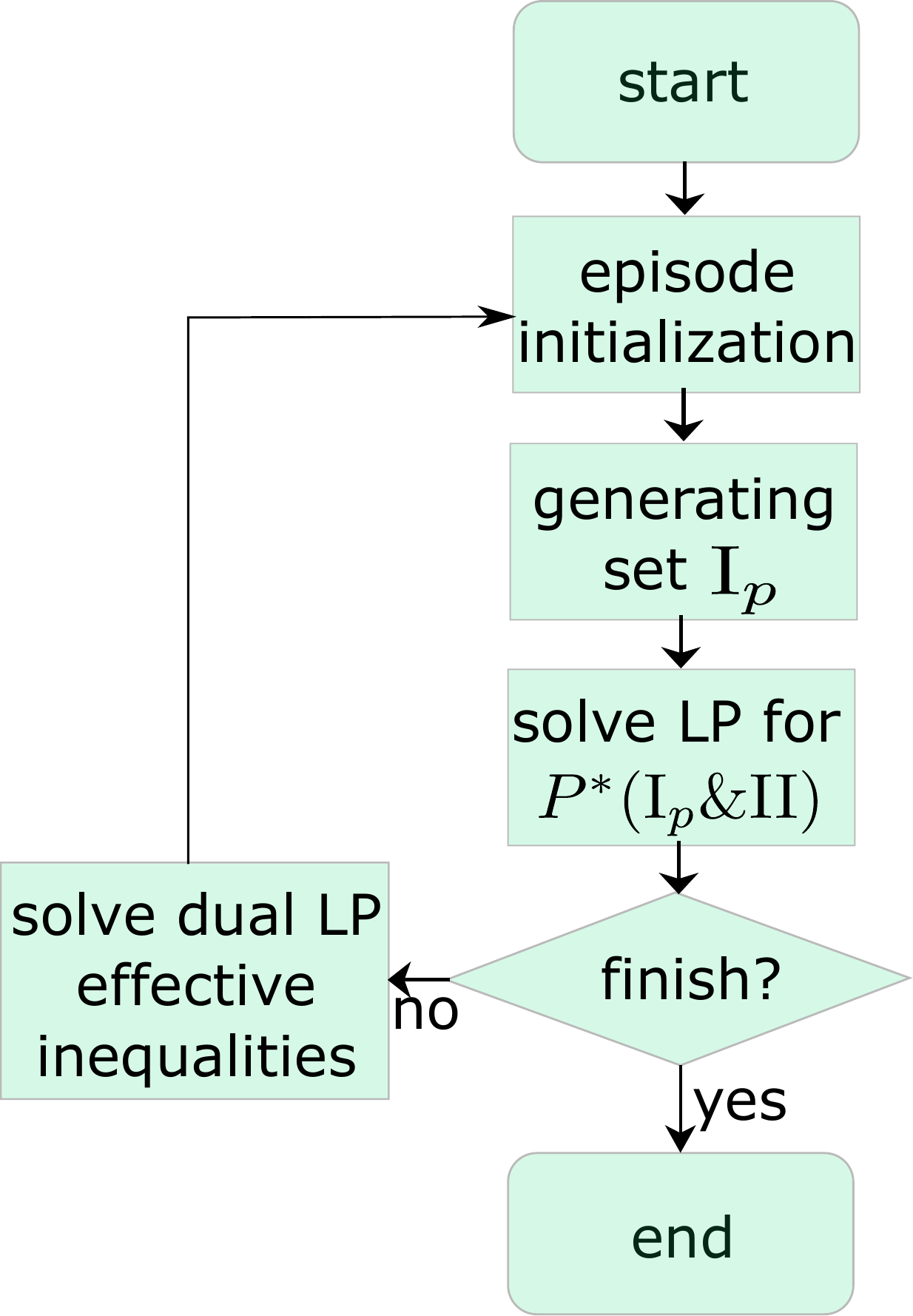}
\caption{The iterative procedure to computer outer bound.\label{fig:RF}}
\end{figure}

Instead of exhaustive enumeration, we need to find the optimal constraint set $\text{I}^*_p$ or a good suboptimal constraint set in a more intelligent manner. To do so, we take an episodic approach, reminiscent to those used in reinforcement learning \cite{sutton2018reinforcement}. The procedure is shown in Fig. \ref{fig:RF}. 

In each episode a set of inequalities $\text{I}_p$ will be chosen, and the corresponding LP solved under the constraints $\text{I}_p\& \text{II}$, which yields an outer bound $P^*(\text{I}_p\&\text{II})$. This bound may be loose, but its value can be viewed as the (delayed) cost for the selected set $\text{I}_p$ in the episode. As shown in \cite{Tian:JSAC13}, the solution to the dual of this LP identifies all the necessary inequalities to prove this outer bound. This implies that the LP solver can in fact also provide the effective inequalities (and the corresponding coefficients) for the outer bound computed in that episode, which can be viewed as evidence on the usefulness of these information inequalities. This learned evidence can be used to initialize in the next episode, from which a new set of inequalties can be generated (potentially in a randomized manner) and the corresponding LP and dual solved again. 

In order to avoid the explosive memory footprint growing exponentially in the number of random variables at run-time, a new data structure is needed to form the LP for larger instances. The approach of pre-assigning an index to each subset of the random variables and then perform symmetry and implication reductions is highly inefficient, not only because this requires too much memory and computation time, but also because this is a huge waste since the eventual LP is quite small and only a tiny portion of the full set of information inequalities is eventually used. Instead we choose to use a key-value data structure to manage the joint entropies and the information inequalities that will be used in the LP. We omit the implementation details due to space constraints.

\subsection{Random $\text{I}_p$ Set Generation}

\label{subsec:localgrowth}

The module of generating $\text{I}_p$ set in the procedure shown in Fig. \ref{fig:RF} is in fact very critical. A naive strategy is to select a random subset in the elemental inequalities, which is however not efficient for the following reason. The majority of the Shannon type inequalities can be written in the following form
\begin{align}
&I(X_{\mathcal{B}};X_{\mathcal{C}}|X_{\mathcal{A}})\notag\\
&=H(X_{\mathcal{A}}\cup X_{\mathcal{B}})+H(X_{\mathcal{A}}\cup X_{\mathcal{C}})\notag\\
&\qquad-H(X_{\mathcal{A}}\cup X_{\mathcal{B}}\cup X_{\mathcal{C}})-H(X_{\mathcal{A}})\geq 0, \label{eqn:MIseparate}
\end{align}
i.e., each such constraint only has $4$ joint entropy terms, despite the fact that there are $2^N-1$ joint entropy terms in the problem, where $N$ is the number of random variables. It follows that if only a small number of information inequalities are selected uniformly at random, most joint entropy terms in them will only appear once. However since the objective function only involves very few quantities (such as the eventual coding rates), all other joint entropy terms in the inequalities must cancel out each other when forming the eventual bound. If each joint entropy term appears only once in the set of selected inequalities, they will not have a counterpart to cancel out with. Therefore, this system of constraints has many useless ones (many inequalities cannot be effective), leading to a weak converse bound. 

Given this observation, we use a different strategy to generate the constraint set $\text{I}_p$. We start from a bootstrapping set of joint entropies, which can be initialized at the start of each episode using learned evidence. For example, it can be the joint entropy terms that are present in the problem-specific constraints, or it can be those found useful in the previous episodes. We then gradually introduce new joint entropy terms and information inequalities: take intersections and union of the terms in the bootstrapping set to form new joint entropy terms and to also add new inequalities.

\begin{figure}[t]
\centering
\includegraphics[width=0.47\textwidth]{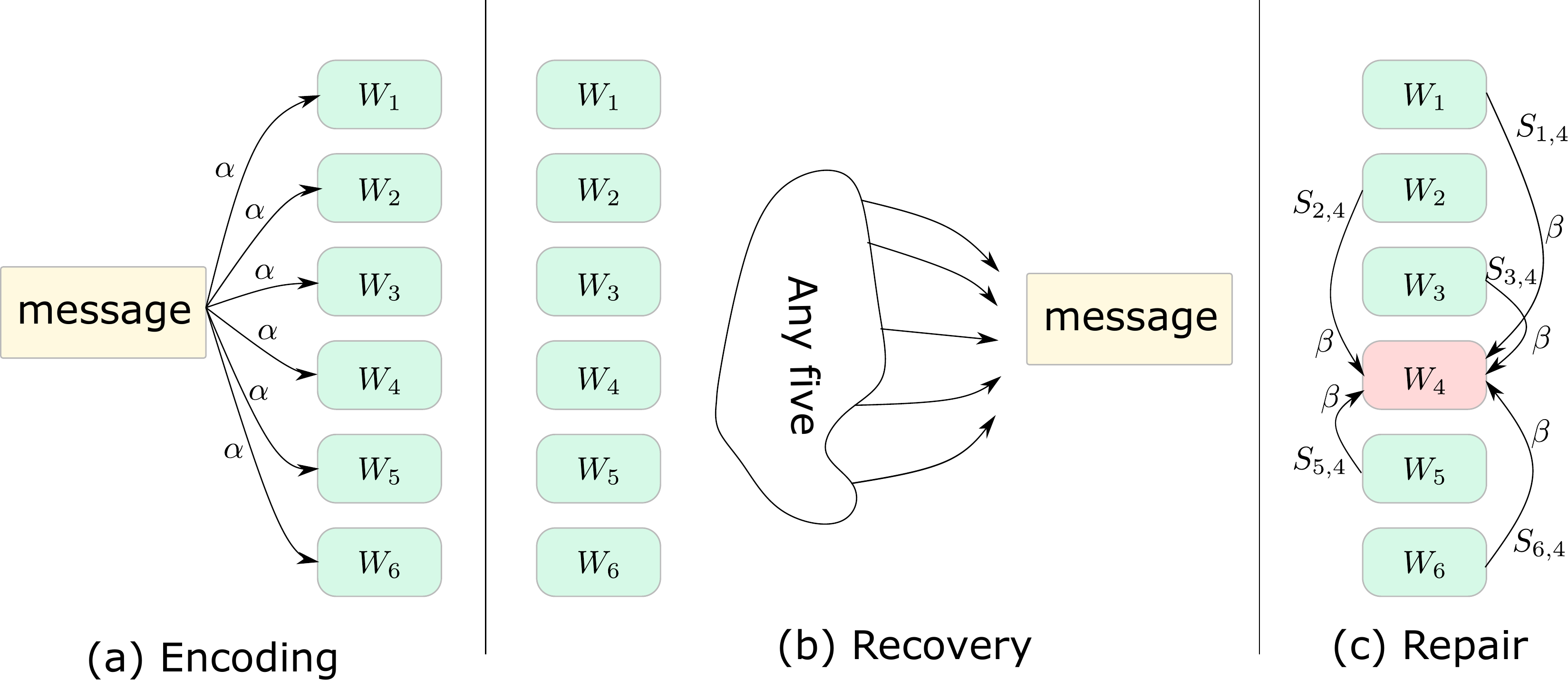}
\caption{The regenerating code problem with $(n,k,d)=(6,5,5)$. \label{fig:rg}.  }
\end{figure}

\subsection{Side Information: Potentially Optimal Code Constructions}

The selection of the set $\text{I}_p$ shown in Fig. \ref{fig:RF} can be made more efficient using different methods based on intuitions that may be mathematically less precise. In this work, we use the following intuition: if a code construction is optimal, i.e., the induced inner bound matches an outer bound derived using a set of information inequalities, then the joint entropy values induced by this code must make all these information inequalities hold with equality. This intuition was used in \cite{tian2017matched,tian2019capacity} to derive certain important properties of optimal codes, and to help construct efficient codes; this approach was termed ``reverse engineering''. In this work, we will instead use potentially optimal code constructions to help narrow down the information inequalities used to form the set $\text{I}_p$.

To be more precise, consider a code construction $\mathbb{C}$ that induces the random variables $X'_1,X'_2,\ldots,X'_N$ and subsequently $2^N-1$ joint entropy values. Suppose we wish to establish and prove a particular outer bound that matches this code construction using the procedure given in Fig. \ref{fig:RF}, then we can restrict the selection of $\text{I}_p$ to the information inequalities that satisfy the following conditions: when $X'_1,X'_2,\ldots,X'_N$ are substituted into them, the inequalities become equality. For example, suppose the construction gives $H(X'_1)=1$, $H(X'_2)=1$, and $H(X'_1,X'_2)=2$, then the elemental inequality
\begin{align}
I(X_1;X_2)\geq 0
\end{align}
indeed holds with equality, thus can be selected into $\text{I}_p$. Otherwise, an inequality will not be allowed into  $\text{I}_p$.

\section{An Application: The Storage-Repair Tradeoff in Regenerating Codes}

In this section, we apply the proposed new computational method on the problem of regenerating codes, and present an outer bound that is tighter than the best known bounds derived analytically. 

\subsection{The Regenerating Code Problem}

The $(n,k,d)$ regenerating code problem  \cite{Dimakis:10,drw:11:snc} deals with the situation that a unit-sized message is stored in a distributed manner in $n$ nodes, each having capacity $\alpha$ 
(Figure~\ref{fig:rg}(a)). Two coding requirements need to be satisfied: 1) the message can be recovered from any $k$ nodes
(Figure~\ref{fig:rg}(b)), 
and 2) any single node can be repaired by downloading $\beta$ amount of information from any $d$ of the other nodes 
(Figure~\ref{fig:rg}(c)).
The fundamental limit of interest is the optimal tradeoff between the storage cost $\alpha$ and the download cost $\beta$. We will consider the case $k=d=n-1$ and particularly $n=6$ in this work. In this setting, the stored contents are $W_1,W_2,W_3,W_4,W_5,W_6$, and the repair message sent from node $i$ to repair $j$ is denoted as $S_{i,j}$. The set of the random variables in the problem are therefore
\begin{gather*}
W_1,W_2,W_3,W_4,W_5,W_6\notag\\
S_{1,2},S_{1,3},S_{1,4},S_{1,5},S_{1,6},\notag\\
S_{2,1},S_{2,3},S_{2,4},S_{2,5},S_{2,6}\notag\\
S_{3,1},S_{3,2},S_{3,4},S_{3,5},S_{3,6}\notag\\
S_{4,1},S_{4,2},S_{4,3},S_{4,5},S_{4,6}\notag\\
S_{5,1},S_{5,2},S_{5,3},S_{5,4},S_{5,6}\notag\\
S_{6,1},S_{6,2},S_{6,3},S_{6,4},S_{6,5}. 
\end{gather*}

It can be observed that the problem has a significant amount of symmetry, which can be used to reduce the scale of the LP. For a given instance with $n$ nodes, the optimization problem in (\ref{eqn:LPmin}-\ref{eqn:contraintII}) has the following objective function
\begin{align}
\text{minimize: }& f_0= \alpha + \eta\beta \label{eqn:LPminRege}
\end{align}
where $\eta$ is a non-negative value which controls the direction of the supporting hyperplane direction of the target converse bound, or it can be viewed as a Lagrangian multiplier for the constraint on $\beta$.
Shannon-type inequalities (I) can be enumerated in a straightforward manner ). The problem-specific inequalities (II) are
\begin{align}
&H(W_i)-H(W_i,\{S_{i,j},j\neq i\})=0, \quad i=1,2,\ldots,n\\
&H(\{S_{i,j},i\neq j\})-H(W_j,\{S_{i,j},i\neq j\})=0, \notag\\
&\qquad\qquad\qquad\qquad\qquad\qquad\qquad j=1,2,\ldots,n\\
&H(W_i)\leq \alpha,\quad i=1,2,\ldots,n\\
&H(S_{i,j})\leq \beta,\quad i=1,2,\ldots,n, j\neq i\\
&H(\{W_i,i\in\mathcal{N}\})\geq 1, \quad \mathcal{N}\subset \{1,2,\ldots,n\}, |\mathcal{N}|=n-1.\label{eqn:contraintIIRege}
\end{align}
The first two constraints are the repair conditions, i.e., the helper messages $\{S_{i,j},j\neq i\}$ can be produced from the message $W_i$, and each stored message $W_j$ can be repaired using the helper messages $\{S_{i,j},i\neq j\}$. The third and fourth constraints bound the storage and the repair message costs. The last inequality guarantees any $n-1$ nodes can reconstruct the original data of unit size. This formulation for the $(4,3,3)$ case was used in \cite{Tian:JSAC13} to yield a tight characterization. 

This representation has a total of $36$ random variables,  which is the standard form used in the literature. However, we can simplify the representation:  the stored content $W_i$ can be omitted and replaced by the repair contents $\{S_{i,j}: j\neq i\}$. This alternative representation has $30$ random variables, and we shall use it in the discussion that follows.

The optimal tradeoff for the case $(4,3,3)$ was established in \cite{Tian:JSAC13}, and the case $(5,4,4)$ in \cite{Tian:15-2}, both through the computational approach. When restricting to linear codes, the optimal $(\alpha,\beta)$ tradeoff for general $(n,k=n-1,d=n-1)$ has indeed been found \cite{duursma2018shortened,prakash2015storage}, however, information theoretic outer bounds without the linear code restriction turn out to be extremely difficult to establish, despite considerable efforts \cite{sasidharan2014improved,duursma2014outer,mohajer2015new,sasidharan2016outer}. 
There does not exist tight outer bounds for $n\geq 6$ in terms of information theoretical optimality, i.e.,  the information theoretic fundamental limits.

\subsection{Potentially Optimal Codes: Canonical Layered Codes}

The inner bound induced by the canonical layered storage code \cite{Tian:15} matches the information theoretic bounds for the $(4,3,3)$ and $(5,4,4)$ cases \cite{Tian:JSAC13,Tian:15-2}, and it also matches the linear code outer bound \cite{duursma2018shortened,prakash2015storage}. Given this result, it has been strongly suspected that this code is in fact also optimal in the information theoretic sense for general $(n,k=n-1,d=n-1)$. The achieved storage-repair tradeoff pairs are
\begin{align}
(\alpha,\beta)=\left(\frac{r}{n(r-1)},\frac{r}{n(n-1)}\right),\quad r=2,3,\ldots,n.
\end{align}

 This code construction can be described as follows. A code parameter $r=2,3,\ldots,n$ is first chosen, and the message consists of a total of $M={n \choose r}(r-1)$ code symbols in a sufficiently large alphabet. Each $(r-1)$ symbols is grouped into a parity group, and a single parity symbol is produced from this parity group. Each parity group is associated with a specific subset of nodes in $\{1,2,\ldots,n\}$ of cardinality $r$, and the message symbols and the parity symbol are placed at the node in this subset of nodes, one symbol at a node. To repair a node, the parity groups involving the failed node are enumerated, and for each such parity group, the symbols other than the one on the failed node are transmitted from the nodes that store them. Given this description, the joint entropy of any given subset of random variables can be calculated simply by counting: first determine how many parity groups are involved, and then for each parity group with a total of $t$ symbols, count it as $\max(t,r-1)$; the final entropy value is the summation over all the involved parity groups, normalized by $M$. 
 
 Consider an example of $r=3$. We then have before the normalization
 \begin{align}
 H(S_{1,2},S_{2,1})=3+3=6,
 \end{align}
where the parity groups of $\{1,2,3\}$, $\{1,2,4\}$, and $\{1,2,5\}$ each contribute two symbols. With this simple counting procedure, we can use the intuition discussed in Section III.D to narrow down the information inequalities.

\subsection{The New Outer Bound for the $(6,5,5)$ Case}

\begin{figure}[t]
\centering
\includegraphics[width=0.48\textwidth]{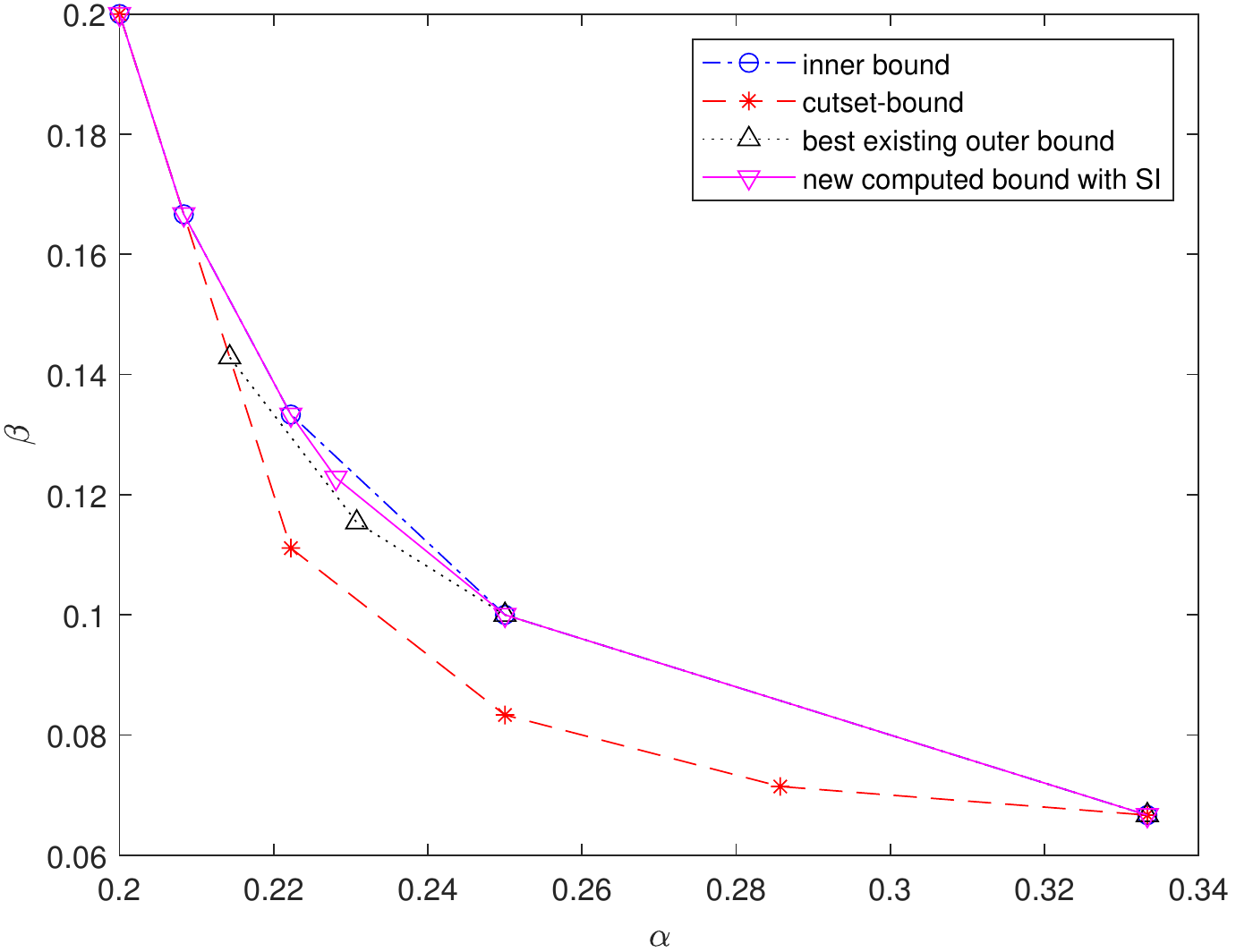}
\caption{Inner and outer bounds for the $(n,k,d)=(6,5,5)$ regenerating code. \label{fig:rg655}.  }
\end{figure}

The new outer bound is illustrated in Fig. \ref{fig:rg655}. For reference, we include the inner bound induced by the canonical layered code \cite{Tian:15}, the cutset bound \cite{Dimakis:10} which is also the functional repair optimal tradeoff, and the best known analytically derived information theoretic bound \cite{mohajer2015new}. It can be seen that the new approach allows us to find tight bounds for all but one segment on the optimal tradeoff. The new bound is considerably tighter than the existing analytically derived bound. 
For the segment that the outer bound does not match the inner bound, the out bound is formed by 
\begin{align*}
27\alpha+15\beta\geq 8,\quad 26\alpha+25\beta\geq 9.
\end{align*}
We note that the bound is not necessarily the best that can be derived using Shannon-type inequalities. 

This result is computed on a workstation with 64GB RAM, and the result was obtained within ten minutes. We note that since the problem has at least 30 random variables, the previous computational approach can not initialize or yield any result on the same hardware platform. Since these bounds are proved by computation, we do not write them in the usual form as a sequence of inequalities; instead, the data for the proof of the new bound can be found at \cite{RGproof}.

 \section{Conclusion}
 
 We propose a new computational approach by converting the entropy LP into a maximin optimization problem. The new formulation naturally leads to an iterative optimization procedure, in which intuitions such as potentially optimal code constructions can be used. As an application of the new approach, we revisit the regenerating code problem with $(n=6,k=5,d=5)$ and obtained new and tighter outer bound. 
 
 There have been more recent work using computational approach based on the entropy linear program approach, e.g., \cite{cao2020characterizing,li2021automated,yeung2021machine}. Some of these approaches can suffer from the same exponential increase of storage and computational complexity when the problem is large, and the proposed approach can potentially be used in them to reduce the computation burden and make progress on seemingly impossible difficult cases. As part of our on-going work, we are studying the application of the proposed approach on larger cases of regenerating code, coded caching, and private information retrieval problems, and developing more efficient methods to incorporate other intuitions in the algorithm.

\bibliographystyle{IEEEtran}
 \newcommand{\noop}[1]{}

\end{document}